\documentclass{JHEP3}

\usepackage{amsmath}
\usepackage{amssymb}
\usepackage{graphicx}

\newcommand{\AP}{\alpha^{\prime}}
\newcommand{\pd}{\partial}

\newcommand{\tit}{\tilde{\tau}}

\newcommand{\diag}{\mathop{\mathrm{diag}}\nolimits}

\newcommand{\Ac}{\mathcal{A}}
\newcommand{\Rc}{\mathcal{R}}

\newcommand{\Gc}{\mathcal{G}}

\newcommand{\Nc}{\mathcal{N}}

\newcommand{\const}{\text{const}}

\title{Cosmological Signature of Tachyon Condensation}

\author{Irina~Ya.~Aref'eva\\
Steklov Mathematical Institute of RAS, Gubkin st. 8, 119991 Moscow,
Russia, E-mail: \email{arefeva@mi.ras.ru}}

\author{Alexey~S.~Koshelev\footnote{On leave from \textit{Steklov
Mathematical Institute of RAS, Gubkin st. 8, 119991 Moscow,
Russia, e-mail:} \texttt{koshelev@mi.ras.ru}}\\
Theoretische Natuurkunde, Vrije Universiteit Brussel and
The International Solvay Institutes,
Pleinlaan 2, B-1050 Brussels, Belgium,
E-mail: \email{koshelev@tena4.vub.ac.be}}

\abstract{We consider the dynamics of the open string tachyon condensation in
a framework of the cubic fermionic String Field Theory including a
non-minimal coupling with closed string massless modes, the graviton
and the dilaton.
Coupling of the open string tachyon and the dilaton is motivated by
the open String Field Theory in a linear dilaton background and the flat
space-time.
We note that the dilaton gravity provides several
restrictions on the tachyon condensation and show explicitly that the influence of the dilaton
on the tachyon condensation  is essential and provides a
significant effect: oscillations of the Hubble
parameter and the state parameter become of a cosmological
scale. We give an estimation for the period of these oscillations
$(0.1-1)$~Gyr and note a good agreement of this period with the
observed oscillations with a period $(0.15-0.65)$~Gyr in a distribution of quasar spectra.
}

\keywords{Cosmology of Theories beyond the SM, String Field Theory}

\preprint{}

\begin{document}


\section{Introduction}

Contemporary cosmological observational data~\cite{data} strongly
support that the present Universe exhibits an accelerated expansion
providing thereby an evidence for a dominating Dark Energy (DE)
component. Recent results of WMAP~\cite{Komatsu} together with the
data on Ia supernovae give the following bounds for the DE state
parameter
$
w_{\text{DE}}=-1^{+0.14}_{-0.11}
$
or without an a priori assumption that the Universe is flat and
together with the data on large-scale structure and supernovae
$w_{\text{DE}}=-1.06^{+0.13}_{-0.08}$.

This range of $w$ includes quintessence models,
$w>-1$~\cite{Wetterich,Peebles}, containing an extra light scalar
field which is not in the Standard Model set of fields~\cite{Yao}, the cosmological constant, $w=-1$~\cite{S-St,Padmanabhan-rev}, and
``phantom'' models, $w<-1$, which can be described by a scalar
field with a ghost (phantom) kinetic term. In this case all natural
energy conditions are violated and there are problems of instability
both at the classical and quantum levels \cite{Caldwell,Woodard}.
Models with a crossing of the $w=-1$ barrier are also a subject of
recent studies. Simplest ones include two scalar fields (one phantom
and one usual field, see~\cite{AKVtwofields,quint} and refs.
therein). General $\kappa$-essence models~\cite{mukhanov,wei} can
have both $w<-1$ and $w\geqslant -1$ but a dynamical transition
between these domains is forbidden under general
assumptions~\cite{Vikman}.

Some projects are directly aimed at exploring whether $w$
varies with the time or is a constant (see \cite{Komatsu} and refs. therein). Varying $w$ obviously
corresponds to a dynamical model of the DE which generally speaking
includes a scalar field. Modified models of General Relativity also
generate an effective scalar field (see for example~\cite{modGR} and
refs. therein). Other DE models based on brane-world scenarios are
presented in~\cite{brane}. A comprehensive review \cite{sami_review}
and references therein may provide the reader with a more detailed
discussion of the DE dynamics.

In the present paper we continue a research along the lines of \cite{Arefeva,AK,K} (see also \cite{J} where numerical methods for studying corresponding nonlinear
and nonlocal models in the Friedman metric were developed) and investigate in
more generality cosmological models coming from the open  String Field Theory (SFT)
tachyon dynamics, namely from the cubic SFT formulation \cite{cubic} of the open
fermionic NSR string with the GSO$-$ sector \cite{NPB} (see \cite{review-sft} for a
review).
The open string tachyon when all massive states are integrated out by means of
equations of motion acquires a non-trivial potential with a non-perturbative minimum.
Rolling of the tachyon from the unstable perturbative extremum towards this minimum
describes, according to the Sen's conjecture \cite{Sen-g}, a transition of an unstable D-brane
to a true vacuum where no perturbative states of the open string are present.
A solution corresponding to a true non-perturbative vacuum in the open fermionic NSR
string with the GSO$-$ sector described by the modified  cubic SFT \cite{NPB} has been
found recently \cite{AGM} and the Sen's conjecture has been proved (for the bosonic string
the Schnabl solution proves the Sen's conjecture \cite{MSchnabl}). The obtained solution
contains the tachyon mode as well as other excitations.
One can, as a reasonable approximation, restrict a number of these excitations and
integrate them out with a result of a Mexican hat potential for the tachyon field.
It happens that there is a rolling solution for the tachyon
(see \cite{yar,AJK} on a numerical construction, about an existence of a solution see
\cite{vladimirov}).

This rolling tachyon solution exhibits under special conditions late time
oscillations with a period of the string scale giving also an oscillating behavior
to the state parameter in the Friedman metric \cite{AK}. A full numeric space-homogeneous solution to this problem in the FRW metric has been found in \cite{J}. One can expect that a nonlinear tachyon
equation describes the dynamics in a transition region whereas a linear approximation
can be taken for the late time behavior.
It is very interesting to clarify the question do other string modes affect
the late time behavior and, if yes, is there a theoretical possibility to make
these oscillations of an observable scale?
Since it is assumed that in the true string vacuum  only the closed strings exist it is
natural first to take into account massless modes of the closed string.
These modes, the graviton, the dilaton, and the anti-symmetric $p$-form field form the gravitational
background\footnote{In the sequel we put the $p$-form equal zero.}.

In this paper we consider a non-minimal coupling of the
open string tachyon and the dilaton motivated by an investigation of the linear dilaton
background in the flat space-time \cite{FMS,AZ,recentSchnabl}.
Analyzing cosmological consequences of our model we demonstrate that
there is a theoretical possibility
to make oscillations in the Hubble parameter and $w$ of cosmological
scales. A period of these oscillations is estimated to be of order $(0.1-1)$~Gyr for natural values of parameters of the model. This number is intriguingly close to the period of oscillations in the $z$-distribution of quasar spectra observed and confirmed recently \cite{varshalovich} (here $z$ is the redshift parameter). The reported value is $(0.15-0.65)$~Gyr.

The paper is organized as follows. In Section~2 we set up the model
and write down the equations of motion. In Section~3 we mention  some facts about
the dilaton gravity and present several exact solutions.
In Section~4 we remind the way one can analyze the dynamics of the tachyon around the non-perturbative vacuum. In Section~5 we demonstrate that dilaton field can give an effect of the Hubble parameter and $w$ oscillations. In the last Section we conclude with a discussion on the obtained results.


\section{Action and Equations of Motion}

We work in $1+3$ dimensions but keep to some extent a general number of dimensions denoted as $D$. The coordinates are denoted by $x^{\mu}$ with Greek indices running
from 0 to $D-1$.

Action motivated by the closed string field theory reads \cite{zw_close-sol,zw_close-sft}
\begin{equation}
S_c=\int
d^Dx\sqrt{-g}\frac{e^{-\Phi}}{2\kappa^2}\left(R+\pd_\mu\Phi\pd^\mu\Phi-g_u{U(\Phi)}-\frac{\pd_\mu
T\pd^\mu T}2+\frac{T^2}{\AP}-\frac1{\AP}{V(\bar T)}\right).
\label{action_modelc}
\end{equation}
Here $\kappa$ is the gravitational coupling constant
$\kappa^2=8\pi G=\frac1{M_P^2}$ (these assignments are true in $D=4$), $G$ is the Newton's constant, $M_P$ is the Planck mass, $\AP$ is the string length squared, $g$ is a metric, \begin{equation*}\bar T=\Gc_c(\AP\Box)T\end{equation*} with
$\Box=D^{\mu}\pd_{\mu}$ and $D_\mu$ being a covariant
derivative, $\Phi$ is the dilaton field and $T$ is the closed string tachyon. Fields are dimensionless while $[g_u]=\text{length}^{-2}$. $\Gc_c$ is supposed to be an analytic function of the argument. $V(\bar T)$ is a closed string tachyon potential. Factor $1/\AP$ in front of the tachyon potential looks unusual and can be easily removed by a rescaling of fields. For our purposes it is more convenient keeping all the fields dimensionless.

Action (\ref{action_modelc}) describes the low-energy dynamics of the metric, the dilaton, and
the closed string tachyon. Kinetic terms as well as potential terms up to fifth
order \cite{KP,zw_close-sft,moeller} can be computed using closed
SFT \cite{BZ-sft} (see \cite{zw_close-sft}
for a relation of sigma model fields and
string fields). We keep higher derivatives and therefore higher order in
$\AP$ terms in the potential.
All these terms do contribute to the dynamics near a nonzero vacuum expectation
value.

$U(\Phi)$ is a dilaton potential introduced for generality. If exists, it would be good if the canonical dilaton property that $\Phi+\const$ results at most in a renormalization of coupling constants is preserved.
Moreover this potential may contain non-local operators acting on the dilaton field. In the rest we assume that $U(\Phi)$ does not depend on the metric keeping such a possibility as an open question for the further study.
Action motivated by the open string field theory reads
\begin{equation}
S_o=\int
d^Dx\sqrt{-g}\frac{1}{g_o^2}\left(e^{-\Phi/2}\left(-\frac{1}2g^{\mu\nu}\pd_\mu
\tau\pd_\nu \tau+\frac1{2\AP} \tau^2\right)-\frac1{\AP}\widehat{e^{-\Phi/2}}v(\tit)\right).
\label{action_premodelo}
\end{equation}
Here $g_o$ is the open string coupling, $[g_o]=\text{length}$ in $D=4$, \begin{equation*}\tit=\Gc_o(\AP\Box)\tau,\end{equation*} hat denotes an operator inverse to the tilde such that \begin{equation*}\hat\varphi=\Gc_d(\AP\Box)\varphi\text{ and }\tilde{\hat{\varphi}}=\varphi,\end{equation*} and $\tau$ is the open string tachyon, it is dimensionless. Both $\Gc_o$ and its inverse are supposed to be analytic functions of the argument. $v(\tit)$ is an open string tachyon potential. A peculiar coupling of the tachyon potential with the dilaton involving an action of a non-local operator hat on the dilaton exponent is supported considering the linear dilaton CFT. Corresponding calculations and further technical details are accumulated in Appendix~A. Also, the above mentioned coupling of the dilaton to the open string tachyon potential may be hidden by virtue of the property
\begin{equation*}
\int
d^Dx\sqrt{-g}A\Box B=\int
d^Dx\sqrt{-g}B\Box A+\text{boundary terms}
\end{equation*}
yielding
\begin{equation}
S_o=\int
d^Dx\sqrt{-g}\frac{e^{-\Phi/2}}{g_o^2}\left(-\frac{1}2g^{\mu\nu}\pd_\mu
\tau\pd_\nu \tau+\frac1{2\AP} \tau^2-\frac1{\AP}\widehat{v(\tit)}\right).
\label{action_modelo}
\end{equation}
Both potentials $V$ and $v$ are supposed to be analytic functions in their arguments.

We study a minimal gravitational coupling of lightest open and closed string modes using the action
\begin{equation}
S=S_c+S_o.
\label{action_model}
\end{equation}
The equations of motion become as follows. For the metric, it reads
\begin{equation}
\begin{split}
&2G_{\mu\nu}+2D_\mu \pd_\nu \Phi+g_{\mu\nu}\left(-2\Box \Phi
+\pd\Phi^2+g_uU\right)-\pd_\mu T\pd_\nu T+g_{\mu\nu}\left(\frac{\pd
T^2}2+\frac{V-T^2}{\AP}\right)+\\&+\frac{2\kappa^2}{g_o^2}e^{\Phi/2}\left(-\pd_\mu \tau\pd_\nu \tau+{g_{\mu\nu}}\left(\frac{\pd
\tau^2}2+\frac{2e^{\Phi/2}\widehat{e^{-\Phi/2}}{v}-\tau^2}{2\AP}\right)\right)+e^\Phi\frac{{\Nc}_{\mu\nu}}{\AP}=0.
\end{split}
\label{EOM_gm}
\end{equation}
Here $G_{\mu\nu}=R_{\mu\nu}-\frac12Rg_{\mu\nu}$ is the Einstein
tensor and $\Nc_{\mu\nu}$ represents
terms coming from a variation of  non-local quantities w.r.t. the
metric. Explicit expressions are presented in Appendix~B, formula (\ref{explicitNc}).
For the field $\Phi$ the equation of motion is
\begin{equation}
\begin{split}
&R+2\Box\Phi-\pd\Phi^2-g_u(U-U')+\left(-\frac12\pd
T^2+\frac{T^2-V}{\AP}\right)+\\
+&\frac{\kappa^2}{g_o^2}e^{\Phi/2}\left(-\frac{1}2\pd
\tau^2+\frac{\tau^2-2\hat{v}}{2\AP} \right)=0.
\end{split}
\label{EOM_phim}
\end{equation}
Prime means a derivative w.r.t. the argument.
For the tachyon fields one has
\begin{eqnarray}
\Box T-g^{\mu\nu}\pd_\mu\Phi\pd_\nu T+\frac{2T}{\AP}-\frac1{\AP}e^\Phi\left(\overline{e^{-\Phi}V'}\right)&=&0,
\label{EOM_Tm}\\
\Box \tau-g^{\mu\nu}\frac{\pd_\mu\Phi\pd_\nu \tau}2+\frac{\tau}{\AP}-\frac1{\AP}e^{\Phi/2}\widetilde{\left(\widehat{e^{-\Phi/2}}v'\right)}&=&0.
\label{EOM_tm}
\end{eqnarray}
We can express $R$ from (\ref{EOM_phim}) and substitute it into (\ref{EOM_gm}) yielding
\begin{equation}
\begin{split}
&2R_{\mu\nu}+2D_\mu \pd_\nu \Phi+g_{\mu\nu}g_uU'-\pd_\mu T\pd_\nu T+\\
&-\frac{2\kappa^2}{g_o^2}e^{\Phi/2}\pd_\mu \tau\pd_\nu \tau+g_{\mu\nu}\frac{\kappa^2}{g_o^2}e^{\Phi/2}\left(\frac{\pd
\tau^2}{2}-\frac{\tau^2}{2\AP}+\frac{2e^{\Phi/2}\widehat{e^{-\Phi/2}}v-\hat{v}}{\AP}\right)+e^\Phi\frac{{\Nc}_{\mu\nu}}{\AP}=0.
\end{split}
\label{EOM_gm2}
\end{equation}
Further, contracting the latter equation with $g_{\mu\nu}$ we can express $R$ again and resubstitute it into (\ref{EOM_phim}) with the following result
\begin{equation}
\begin{split}
&\Box\Phi-\pd\Phi^2-g_uU+g_u\left(1-\frac D2\right)U'+\frac{T^2-V}{\AP}-e^\Phi\frac{g^{\mu\nu}\Nc_{\mu\nu}}{2\AP}+\\&+
\frac{\kappa^2}{g_o^2}e^{\Phi/2}\frac{\tau^2-2\hat{v}}{2\AP}-
\frac D2\frac{\kappa^2}{g_o^2}e^{\Phi/2}\left(\frac{\pd
\tau^2}{2}-\frac{\tau^2}{2\AP}+\frac{2e^{\Phi/2}\widehat{e^{-\Phi/2}}v-\hat{v}}{\AP}\right)=0.
\end{split}
\label{EOM_phim2}
\end{equation}
Equations (\ref{EOM_Tm})-(\ref{EOM_phim2}) give full (redundant) set of equations to be solved. At least one equation can be excluded by virtue of the Bianchi identity $D^\mu G_{\mu\nu}\equiv0$.

From now on the only metric we will be interested in is the spatially flat FRW metric of the form $g_{\mu\nu}=\diag(-1,a^2, a^2, a^2)$ where $a(t)$ is the space-homogeneous scale factor, $t$ is the cosmic time. Also all the equations in the sequel are formulated for $D=4$. The Hubble parameter is defined as $H=\dot a/a$ where dot is the time derivative. Also all other fields will be taken space-homogeneous. Some useful quantities in this metric are as follows
\begin{equation*}
\Gamma_{ij}^0=Hg_{ij},~\Box=-\pd_t^2-3H\pd_t,~R_{\mu\nu}=\left(\begin{array}{cc}-3(\dot H+H^2)&0\\0&g_{ij}(\dot H+3H^2)\end{array}\right),~R=6\dot H+12H^2,
\end{equation*}
where we omitted space derivatives in the operator $\Box$.

A distinguished feature of our model is that the tachyon field is accompanied by an
action of a non-local operator.\footnote{Analogous models come from the $p$-adic
formulation~\cite{BFOW,frampton,padic} of the tachyon dynamics.} This non-locality breaks
constraints on a phantom divide crossing making possible a dynamical transition between
quintessence and phantom phases. It is explicitly shown in \cite{AK} that a non-locality
may provide a periodic crossing of the $w=-1$ barrier. Being a string theory limit the
model addresses all stability issues \cite{SD,AV1} to the string theory. Cosmological
applications of such models initially were studied in \cite{AKV,AV,AK}. Also the dynamics
of a non-local tachyon on a cosmological background is studied in \cite{calcagni}. In
\cite{biswas} and refs. therein inflationary models and their relation to non-gaussianities are studied using a similar non-local Lagrangian.


\section{Vacuum}
\label{sec:vac}

We have an expectation that starting from the perturbative unstable tachyonic vacuum system evolves towards its true ground state. We are aimed at the analysis of the system dynamics around its true vacuum but first it is interesting to analyze what the vacuum is. One expects that the tachyon fields rest in the vacuum at the end of their evolution. Rolling of the open string tachyon is attributed to the unstable D-brane decay.
When we consider the NS string and assume that the supersymmetry
is restored in the true vacuum, as a consequence there should not be the closed string tachyon mode.
The bosonic string tachyon motion has been studied in
\cite{zw_close-sol,zw_close-sft,moeller}. We keep however the closed string tachyon in action
(\ref{action_modelc}) for a generality.

In our previous consideration the dilaton field
has not been taken into account.  One of the possibilities is to think that the
dilaton field has been frozen in the early stage of the evolution. Here we are going to explore a more involved
situation when the dilaton contributes to the late time evolution.
Earlier the closed string tachyon condensation including its interaction with the dilaton was studied, for example, in \cite{KP}.
Regarding the dilaton gravity itself we refer the reader to the review paper \cite{gasperini} and refs. therein where the
dilaton gravity and its phenomenology are described in many details. In papers \cite{gasp2,tsujikawa} the dilaton as a string motivated Dark Energy model is examined and several exact solutions are presented. See also \cite{greek} on how the dilaton helps reconstructing the cosmic history. Several versions of a nonlocal dilaton gravity also have been studied in \cite{Ven,AJV}.

Let us note note that provided there is no the dilaton field from the beginning then one has to use equation (\ref{EOM_gm}) with $\Phi=0$ while equations (\ref{EOM_gm2}) and (\ref{EOM_phim2}) are not in the game. Taking vacuum values $T_0$ and $\tau_0$ for the tachyons we see from equation (\ref{EOM_gm}) that it is reduced to
\begin{equation*}
\begin{split}
&G_{\mu\nu}=-g_{\mu\nu}\frac\Lambda2
\end{split}
\end{equation*}
where $\Lambda$ is whatever has left from potential terms of the tachyons, $[\Lambda]=\text{length}^{-2}$. $\Lambda$ may be zero as well. However it is an obvious possibility to build a realistic de Sitter gravitational background since $\Lambda$ is manifestly a cosmological constant. We get a relation $H^2=\frac\Lambda6$
and one redirects questions about an appearance and a smallness of $\Lambda$ to a structure of tachyon potentials.

Switching on the dilaton one has to account its equation of motion. Solutions for the dilaton gravity in stringy context were widely analyzed before (see, for instance, \cite{KP,tsujikawa}). In the case both tachyons rest in their vacua one has
\begin{eqnarray}
2R_{\mu\nu}+2D_\mu \pd_\nu \Phi+g_{\mu\nu}g_uU'+g_{\mu\nu}\frac{\kappa^2}{g_o^2}e^{\Phi/2}\frac{2v-\tau^2}{2\AP}&=&0,
\label{EOM_gm2_0}\\
\Box\Phi-\pd\Phi^2-g_uU+g_u\left(1-\frac D2\right)U'+\frac{T^2-V}{\AP}+
\left(1+\frac D2\right)\frac{\kappa^2}{g_o^2}e^{\Phi/2}\frac{\tau^2-2\hat{v}}{2\AP}&=&0.
\label{EOM_phim2_0}
\end{eqnarray}
In the most important for our model scenario with a linear dilaton $\Phi=-2V_0t$ and $D=4$ the above two equations become
\begin{eqnarray}
3\dot H+3H^2+\frac{g_uU'}2+\Lambda_2e^{\Phi/2}=0,\quad
\dot H + 3H^2+2HV_0+\frac{g_uU'}2+\Lambda_2e^{\Phi/2}&=&0,\label{EOM_gm4vUH12_0}\\
6HV_0+4V_0^2-g_u(U+U')+\Lambda_1-6\Lambda_2e^{\Phi/2}&=&0.
\label{EOM_phim4vUH_0}
\end{eqnarray}
where $\Lambda_1=\frac{T^2-V}{\AP}$ and $\Lambda_2=\frac{\kappa^2}{2g_o^2}\frac{\tau^2-2\hat{v}}{2\AP}$. One immediately sees that in this case Hubble parameter satisfies
$
\dot H=HV_0
$
with a solution
\begin{eqnarray}
H=H_0e^{V_0t}=H_0e^{-\Phi/2}
\label{Hubble0}
\end{eqnarray}
Then one can integrate any of equations (\ref{EOM_gm4vUH12_0}) w.r.t. $\Phi$ in order to get the potential $U$. The result is
\begin{eqnarray}
\frac{g_u}2(U+U_0)=3H_0^2e^{-\Phi}+6H_0V_0e^{-\Phi/2}-2\Lambda_2e^{\Phi/2}
\label{U0}
\end{eqnarray}
accompanied by the consistency condition $4V_0^2+\Lambda_1=U_0$.

The change of $H$ is very slow because $V_0$ should be small according to the rate of change of the Newton's constant. The Hubble parameter on this solution is either grows or vanishes exponentially but a ratio $\dot H/H$ is very small since it is of the same order as the value $\dot G/G$ \cite{GNchange}.


\section{Tachyons Around Vacuum without Dilaton Field}

Before studying the influence of the dilaton on the late time tachyon
dynamics and therefore on the late time cosmology we remind few already known facts about the late
time tachyon dynamics without the dilaton field.
We consider the tachyons near their true vacuum and linearize equations of motion using $T=T_0+Z$ and $\tau=\tau_0+\zeta.$ Let us focus on the open string tachyon field $\tau$. A linearized equation becomes
\begin{equation}
\Box \zeta-g^{\mu\nu}\frac{\pd_\mu\Phi\pd_\nu \zeta}2+\frac{\zeta}{\AP}-\frac{v''(\tau_0)}{\AP}e^{\Phi/2}\widetilde{\left(\widehat{e^{-\Phi/2}}\widetilde{\zeta}\right)}=0.
\label{EOM_tmvac}
\end{equation}
If the dilaton field is a constant (or at least one can neglect its variation compared with the tachyon field) one gets the following equation
\begin{equation}
\Box \widehat{\widehat{\zeta}}+\frac{\widehat{\widehat{\zeta}}}{\AP}-\frac{v''(\tau_0)}{\AP}\zeta=0.
\label{EOM_tmvaczerophi}
\end{equation}
A way of solving such an equation was developed in \cite{AK,K,AJV}. The main idea is that an eigenfunction of the $\Box$ operator with an eigenvalue $\omega^2=\bar\omega^2/\AP$ solves the latter equation if $\bar\omega^2$ solves the transcendental equation
\begin{equation}
\left(\bar\omega^2+1\right) \Gc_o^{-2}(\bar\omega^2)={v''(\tau_0)}.
\label{EOM_tmvaczerophi_character}
\end{equation}
Using a level truncated cubic fermionic string field theory we have $\Gc_o^{-2}(\bar\omega^2)=e^{\beta\bar\omega^2}$. A solution for $\bar\omega^2$ becomes
\begin{equation}
\bar\omega^2=\frac{W(\gamma)-\beta}{\beta},\quad \gamma=v''(\tau_0)\beta e^\beta,
\label{sol_omega2}
\end{equation}
where $W$ is Lambert $W$-function which solves an equation $xe^x=y$ w.r.t. $x$. All $\bar\omega$, $\beta$ and $\gamma$ are dimensionless. Notice, that for $v''(\tau_0)=1$ we have $\omega^2=0$ irrespectively of $\beta$.

In fact there are infinitely many branches of this function in a general situation. Zero branch is distinguished since it gives real values on the positive semi-axis. On the negative semi-axis a branch point of zero and $-1$ branches is located at $-1/e$ and $W_{(0)}(-1/e)=-1$ (hereafter a subscript $(n)$ denotes the branch). For positive $\epsilon$ one finds that $W_{(0)}(-1/e+\epsilon)$ is real while $W_{(0)}(-1/e-\epsilon)$ is complex. In case of complex $\omega^2$ one readily finds that $W_{(-1)}(-1/e-\epsilon)=W_{(0)}(-1/e-\epsilon)^*$. This provides a chance to write a real solution for $\tau$ as a linear combination of functions corresponding to $\omega_{(0)}^2$ and $\omega_{(-1)}^2$. Zero and $-1$ branches give smallest in absolute value
  roots and should be considered as a main contribution to a solution of our linearized equation.

Thus, if $\gamma\geq-1/e$ then $\omega^2$ is real and is given by the zero branch of the Lambert function. A solution to equation (\ref{EOM_tmvaczerophi}) is given by a function which satisfies
\begin{equation}
\Box \zeta=\omega^2\zeta.
\label{EOM_tmvaczerophireal}
\end{equation}
If $\gamma<-1/e$ then $\omega^2$ is complex and is given by the zero and $-1$ branches of the Lambert function. A solution to equation (\ref{EOM_tmvaczerophi}) is given by a linear combination of functions which satisfy
\begin{equation}
\Box \zeta_{(0)}=\omega_{(0)}^2\zeta_{(0)},\quad \Box \zeta_{(-1)}=\omega_{(-1)}^2\zeta_{(-1)}.
\label{EOM_tmvaczerophicomplex}
\end{equation}
Integration constants of solutions $\tau_{(0)}$ and $\tau_{(-1)}$ can be adjusted such that resulting $\zeta=\zeta_{(0)}+\zeta_{(-1)}$ is real. Alternatively, one can say that fluctuations are effectively described by two freely propagating fields with masses squared $-\omega_{(0)}^2$ and $-\omega_{(-1)}^2$. Although a quantity corresponding to the initial field $\zeta$ can be made real both effective scalar fields have complex masses squared. One can consider real linear combinations of the effective fields. In this case one field will be either a ghost or a tachyon. This is in agreement with the Ostrogradski analysis \cite{ostrogradski} (see \cite{AV1,AK,K,AJV} for more details).

Thus, in order to analyze the dynamics of the tachyon at large times one has to solve equation (\ref{EOM_tmvaczerophireal}) (or (\ref{EOM_tmvaczerophicomplex}))
which for a constant Hubble parameter gives as a solution
\begin{equation}
\zeta=e^{-\frac32Ht}\text{Re}\left(\zeta_+e^{\frac12t\sqrt{9H^2-4\omega^2}}+\zeta_-e^{-\frac12t\sqrt{9H^2-4\omega^2}}\right).
\label{sol_tmvaczerophiconstH}
\end{equation}
Recall that $\omega^2$ is given by equation (\ref{sol_omega2}) and is in general a complex number.

Having sketched the main line of solving linear non-local equations we are now ready to attack the question of the influence of the dilaton on the tachyon dynamics and, what is more interesting, a back-reaction on the gravity.


\section{Tachyons Around Vacuum in Dilaton Background}

Now we turn to the main question. What is the influence of the dilaton
on the late tachyon cosmology? We have demonstrated in Section~\ref{sec:vac} that the dilaton gravity
only with special dilaton potentials can give a realistic present time cosmological evolution. Studying the influence of the dilaton field on the tachyon condensation we assume one of these potentials appears in the action.
It would be interesting to get a dilaton
potential directly from the closed string field theory. About the first steps in this
direction see \cite{zw_close-sft,moeller}.
Scales related to the dilaton should be of cosmological magnitudes.
For instance, the present Hubble parameter is related to the Planck mass as $H_{\text{our}}\approx10^{-60}M_P$. The same level of smallness is applicable to a rate of change of the dilaton field. Indeed, in the dilaton gravity $\dot\Phi$ determines a speed of change of the Newton's constant. Thus $\dot\Phi$ should be extremely small. For instance, assuming that Newton's constant changes at most $e$ times during the life-time of the universe we conclude that in a model with a linear dilaton one should have $\Phi\approx\pm H_{\text{our}}t$. It looks counterintuitive that such small quantities can affect somehow processes related to the tachyon condensation. The reason is that all ingredients related to string excitations have $\AP$ as a scale. $\AP$ is the string length squared and can be written as $1/M_s^2$ where $M_s$ is the string mass. This string mass in any case cannot be less than 1~TeV to be compatible with present experiments. Moreover, $M_s$ is often associated with the Planck mass $M_P$.

Examining more carefully equation (\ref{EOM_tmvac}) and the succeeding formulae we observe, however, that parameters $\gamma$ and $\beta$ introduced in the previous Section play a crucial role and new interesting phenomena can emerge. Let us make an approximation which produces a very useful toy equation
\begin{equation}
\Box \zeta+\frac{\zeta}{\AP}-(1+\epsilon) \frac{v''(\tau_0)}{\AP}\widetilde{\widetilde{\zeta}}=0.
\label{EOM_tmvactoy}
\end{equation}
Compared to equation (\ref{EOM_tmvac}) we have dropped a term proportional to the speed of the dilaton\footnote{Strictly speaking this is done to make an analytic analysis possible. Further one can demonstrate numerically that restoring this term does not change the main effect described in this Section.} and assumed that exponents of the dilaton with all non-local operators acting on them can be accounted as a constant $\epsilon$. This latter constant is positive for non-local operators coming from the SFT (see Appendix~A).
Similar to (\ref{sol_omega2})
\begin{equation}
\omega_\epsilon^2=\frac{W((1+\epsilon)\gamma)-\beta}{\beta\AP},\quad \gamma=v''(\tau_0)\beta e^\beta\label{sol_omega2epsilon}
\end{equation}
and we see that we have shifted the parameter $\gamma$ and consequently $\omega^2$. For a linear dilaton $\Phi=-2V_0 t$ this shift $\epsilon$ can be computed in the SFT to be of order $\AP V_0^2$, $[V_0]=\text{length}^{-1}$. Accounting the above discussion about cosmological scales we see that $\epsilon\sim\AP H^2$. For these scales a constant $H$ approximation is easily justified and the dynamics of a linearized tachyon $\tau=\tau_0+\zeta_\epsilon$ is given by an analog of (\ref{sol_tmvaczerophiconstH})
\begin{equation}
\zeta_\epsilon=e^{-\frac32Ht}\text{Re}\left(\zeta_+e^{\frac12t\sqrt{9H^2-4\omega_\epsilon^2}}+\zeta_-e^{-\frac12t\sqrt{9H^2-4\omega_\epsilon^2}}\right).
\label{sol_tmvacepsphiconstH}
\end{equation}

Unshifted $\omega^2$ is entirely determined by the open SFT parameters $\beta$ and $\gamma$. It is very difficult to imagine that stringy processes may give as a final state somewhat with scales of the present day cosmology. Corresponding fine-tuning should be of an extreme precision. If $\omega^2$ is finite then most likely it is finite in units of $\AP$. Such an $\omega^2$ dominates under the square root in (\ref{sol_omega2}) and oscillations become very frequent, and the suppression factor becomes huge in cosmological scales.
A shift of order $\epsilon\sim\AP H^2$ will not change anything.

A distinguished case is $\omega^2=0$ which may have a signature in the late cosmology. To have $\omega^2=0$ one has to require $v''(\tau_0)=1$ irrespectively of $\beta$. This means that the full tachyon potential including the mass term should have zero second derivative in the minimum\footnote{One can check that for this purpose one has to have at least three terms of a different degree greater than 2 of $\tit$ in $v$ provided $v$ is polynomial. It is natural to expect this kind of a potential from the SFT when more mass levels are included in the calculation of the tachyon potential.}. In other words the tachyon should become a long-range field in the true minimum. Parameter $\gamma$ becomes
\begin{equation}
\gamma=\beta e^\beta,\quad \gamma\geq-\frac1e\text{ for real }\beta,\quad \label{realgamma} \gamma=-\frac1e\text{ for }\beta=-1.
\end{equation}
In this case of $\omega^2=0$ (i.e. $v''(\tau_0)=1$) the dilaton which provides $\epsilon\sim \AP H^2$ can change the behavior drastically. Indeed, the expression $9H^2-4\omega_\epsilon^2$ under the square root can become negative. Moreover, if parameter $\beta=-1$ then a positive $\epsilon$ makes $\omega^2$ complex at once according to the properties of the Lambert function as explained in the previous Section. This scenario produces oscillations in $\zeta$ with a frequency that can be observed at cosmological scales.

Corrections to a constant $H$ emerge and these correction can be computed using the technique elaborated in \cite{AK,K}. One can use the corresponding analysis now with an inclusion of the dilaton field which does not complicate calculations. Here we just briefly sketch what happens with a dynamical tachyon. Now we have a time-dependent kinetic term and the potential of the tachyon as well as some impact from non-local terms $\Nc_{\mu\nu}$ in equation (\ref{EOM_gm2}). Once we got an oscillating tachyon these oscillations are translated to the dilaton and the Hubble parameter. Taking a solution for the Hubble parameter and the dilaton and making a linearization around it\footnote{It was mentioned in \cite{K} that one should linearize the scale factor $a(t)$ of course, and not $H$.} one will find corrections which obviously will have oscillations (a frequency and a suppression factor will be doubled compared to the tachyon field). For instance, solution (\ref{Hubble0}) corresponding to the linear dilaton gravity with a specific non-perturbative dilaton potential (\ref{U0}) can be taken. $H$ is not a constant on this solution but $\dot H/H$ is of the same order as $\dot G/G$ \cite{GNchange}. Thus it is valid to take a constant $H$ approximation. Further, using a relation
\begin{equation*}
w=-1-\frac{2\dot H}{3H^2}
\end{equation*}
one finds that the equation of state parameter should be also an oscillating quantity.

To get an insight in numbers we derive let us proceed with a linear dilaton $\Phi=-2V_0 t$. We take $V_0=\sigma H_\text{our}$ meaning that during the evolution of the universe the Newton's constant has changed $e^{2\sigma}$ times, $\sigma$ is dimensionless. This gives a shift $\epsilon=\sigma^2\AP H_{\text{our}}^2=\sigma^210^{-120}M_P^2/M_s^2=\sigma^210^{-120+2n}$ where we put $M_s=10^{-n}M_P$ with $0\leq n\leq16$. The upper bound on $n$ gives strings of TeV mass which is a minimal string mass compatible with current experiments.

Passing from (\ref{sol_omega2}) to (\ref{sol_omega2epsilon}) under our assumption that the full tachyon potential has zero second derivative in the non-perturbative minimum, i.e. $v''(\tau_0)=1$, two significantly different situations appear: $\gamma>-1/e$ and $\gamma=-1/e$.

The most interesting one is $\gamma>-1/e$, i.e. $\beta\neq-1$ (see (\ref{realgamma})). In this case
\begin{equation*}
\omega_\epsilon^2\approx\frac{\epsilon}{(1+\beta)\AP}
\end{equation*}
and for the estimated $\epsilon\sim\sigma^2\AP H_\text{our}^2$ solution (\ref{sol_tmvacepsphiconstH}) becomes
\begin{equation*}
\zeta_\epsilon\approx e^{-\frac32H_\text{our}t}\text{Re}\left(\zeta_+e^{\frac12H_\text{our}t\sqrt{9-4\frac{\sigma^2}{1+\beta}}}+\zeta_-
e^{-\frac12H_\text{our}t\sqrt{9-4\frac{\sigma^2}{1+\beta}}}\right).
\end{equation*}
Negative expression under the square root immediately gives oscillations. In the open SFT truncated up to level zero $\beta$ becomes $-0.523$. This means that oscillations come out of the latter solution if $\sigma>1.036$. Taking $\sigma=1.1$ we get the frequency of oscillations in $\zeta_\epsilon$ is approximately $H_\text{our}/2$. As it was mentioned, this frequency is doubled and is equal to $H_\text{our}$ in the Hubble parameter and the state parameter. This produces oscillations with a period of order 10~Gyr which is of order of the universe age. Taking other value of $\beta$ which can result from an inclusion of the higher massive modes in the SFT analysis we may get smaller values of the period of the oscillations.
Taking $\beta=-0.95$ and $\sigma=1.1$ the period of oscillations of the Hubble parameter and the state parameter becomes of order 1~Gyr. The closer $\beta$ to $-1$ from above the shorter period is. The higher the dilaton speed the shorter period is as well.
Thus we see that in the case $\gamma>-1/e$ one can have an effect of oscillations in the Hubble parameter and the state parameter $w$ accessible for observations.

In the case $\gamma=-1/e$ (i.e. $\beta=-1$) one has $\AP\omega_\epsilon^2\approx-{I\sqrt{2\epsilon}}$, this term most likely dominates under the square root in (\ref{sol_tmvacepsphiconstH}) and the resulting frequency becomes of order
$\sqrt{\sigma}10^{+30-n/2}H_{\text{our}}$. In order to make the period of oscillations of order 1~Gyr we have to assume $\sigma\sim10^{-46}$ for the TeV string mass with $n=16$. This means that the dilaton changes very slowly, but it is difficult to figure out an appearance of a new small parameter $V_0\sim10^{-106}M_P$.

The closed string tachyon can be considered analogously and will give the same effect provided its behavior is similar to the open string tachyon. The problem is that a closed string tachyon action is difficult to compute. The bosonic closed SFT has only a non-polynomial formulation so far. Even a static potential being computed in the bosonic closed SFT including few low mass levels \cite{moeller} does not give a clear understanding of dynamical processes associated with the tachyon motion. At this stage we postpone a detailed investigation of the influence of the closed string tachyon dynamics on the cosmology by noting that it is consistent to put the tachyon field value to a constant.


\section{Summary and Discussion}

We have considered a very interesting interplay of stringy and cosmological processes. It turns out that the open string tachyon condensation may lead to cosmologically significant effects. In short, provided that the full tachyon potential (which is the mass term plus $v$) has zero second derivative in the non-perturbative minimum the late time dynamics of just the tachyon in the FRW background with a constant Hubble parameter is given by (\ref{sol_tmvaczerophiconstH}) with $\omega^2=0$. This means that it is either constant or vanishes monotonically. Second parameter, $\beta$, does not affect this behavior at all. Having coupled to the dilaton the tachyon dynamics changes drastically. $v''(\tau_0)$ is effectively shifted from unit and this produces an oscillating behavior out of solution (\ref{sol_tmvacepsphiconstH}). Now parameter $\beta$ is very important and practically it determines the period and the suppression factor of oscillations. If $\beta\neq-1$ then a period of oscillations can be observed. For $\beta=-0.95$ one gets the period of order 1~Gyr. This is in the case that the dilaton speed is of order $H$. The closer $\beta$ to $-1$ from above the shorter period is. The higher the dilaton speed the shorter period is as well. Being accounted in Friedmann equations these oscillations of the tachyon field produce oscillations in $H$ and the state parameter $w$. We note a good agreement of the estimated  period with the observed oscillations in the $z$-distribution of quasar spectra \cite{varshalovich}. The reported value is $(0.15-0.65)$~Gyr.

If $\beta$ is strictly $-1$ than there is a jump of about 30 orders of magnitude due to analytic properties of Lambert $W$ function. All oscillations become very frequent and their suppression factor is huge. A new bizarre parameter which is the speed of dilaton of order $V_0\sim10^{-106}M_P$ is needed to smooth the situation.

Thus $\beta\neq-1$ looks more plausible since no new strange parameters are required. Corresponding oscillations are added up to a background produced by the dilaton and analyzed in Section~\ref{sec:vac}. For instance, solution (\ref{Hubble0}) corresponding to the linear dilaton gravity with a specific non-perturbative dilaton potential (\ref{U0}) can be taken. $H$ is not a constant on this solution but $\dot H/H$ is of the same order as $\dot G/G$ \cite{GNchange}. Presence of a linear dilaton in this solution is important for the estimations we have made.

The closed string tachyon dynamics can also be considered in a similar fashion and will give similar effect as the open string tachyon. We, however, do not consider it explicitly because of a non clear nature and meaning of a closed string tachyon potential and a vacua structure. Moreover, a formulation of the fermionic closed SFT is not yet known. For our purposes it was consistent to assume a constant closed string tachyon field. We address a consideration of the closed string tachyon dynamics as a very interesting and rich question for the further study.

Coupling of open and closed string modes is another big puzzle worth a deeper analysis. We have taken a model coupling of the dilaton and the open string tachyon briefly argued in Appendix~A. This, however, most likely is not the end of the story and a better understanding of an open-closed strings interaction is of a great importance here. Inclusion of the massless vector field and a relation to the Vacuum SFT are two more open questions answering which may provide a better understanding of the string dynamics.

Searching for a full solution is an interesting but seems to be a non-realistic task. Numeric methods, however, may help one to get an insight in how the full solution behaves. Also numerics will be useful to perform a check of the stability of found solutions.


\section*{Acknowledgements}
The authors are grateful to B.Dragovich, N.Moeller, K.Postnov, M.Schnabl, S.Vernov, B.Zwiebach for useful comments and
discussions. The work is supported in part by RFBR grant
08-01-00798, INTAS grant 03-51-6346 and Russian President's grant
NSh-2052.2003.1. A.K. is supported in part by the Belgian Federal Science Policy Office through the Interuniversity Attraction Poles IAP VI/11, by the European Commission FP6 RTN
programme MRTN-CT-2004-005104 and by FWO-Vlaanderen through project G.0428.06.


\appendix

\section{SFT in Linear Dilaton Background}

Here we do not explain anything related to CFT and SFT calculations referring reader to \cite{FMS,review-sft}.

The worldsheet action for the open string in the linear dilaton background in the flat space-time is given by
\begin{equation*}
S=\frac1{4\pi\AP}\int d\tau d\sigma\sqrt{h}\left(h^{\tau\sigma}\pd_\tau
X^{\mu}\pd_\sigma X_{\mu}+\AP\Rc V_\mu\pd X^\mu+\text{fermions}+\text{ghosts}\right).
\end{equation*}
Here $\Rc$ is the two-dimensional curvature and $V_\mu$ comes from the linear dilaton $\Phi=2V_\mu X^\mu$ playing a role of the background charge. As compared to the ordinary CFT the bosonic stress tensor and the matter supercurrent get modified and look like
\begin{equation*}
T^X=-\frac 1{\AP}:\pd X \pd X:+V\pd^2X,\quad G=i\sqrt{\frac{\AP}{2}}(:\psi \pd X:-\AP V\pd \psi).
\end{equation*}
The anomaly charge is $c^X=D+6\AP V^2$. OPE of two left-moving $X$ is as usual and is given by
\begin{equation*}
X(z)^\mu X(w)^\nu=-\frac{\AP}2\eta^{\mu\nu}\log(z-w)
\end{equation*}
whereas OPE of several exponents is changed as follows
\begin{equation*}
\prod_j e^{2ik_jX(z_j)}=(2\pi)^D\delta\left(\sum_ik_i+iV\right)\prod_{m<n}(z_m-z_n)^{2\AP k_m
k_n}.
\end{equation*}
$e^{2ikX(z)}$ is still a primary operator but with a weight $\AP(k^2+ikV)$. Also a transformation law for the field $X$ is changed to be
\begin{equation*}
f\circ X_{\mu}(z)=X^{\mu}(f(z))+\AP V^\mu\log f'.
\end{equation*}
The BRST charge looks as usual in terms of stress tensors and supercurrents
\begin{equation*}
Q=\oint\frac{dz}{2\pi
i}j_{BRST}=\oint\frac{dz}{2\pi
i}\left(:c\left(T^X+T^{\Psi}+T^{\xi\eta}+T^{\varphi}\right)+bc\pd
c-b\eta\pd\eta e^{2\varphi}+\eta e^{\varphi}G:\right)
\end{equation*}
but new $T^X$ and $G$ should be used. Nilpotency of $Q$ dictates
\begin{equation*}
D+4\AP V^2-10=0.
\end{equation*}

Now we calculate the cubic  action of the fermionic SFT on a string field truncated to weight zero. The action is
\begin{equation*}
\begin{split}
S=\frac1{g_o^2}&\left(\frac1{2\AP}\langle\!\langle
Y_{-2}|\Ac_+,Q\Ac_+\rangle\!\rangle+\frac13\langle\!\langle
Y_{-2}|\Ac_+,\Ac_+,\Ac_+\rangle\!\rangle+\right.\\
&+\left.\frac1{2\AP}\langle\!\langle
Y_{-2}|\Ac_-,Q\Ac_-\rangle\!\rangle+\langle\!\langle
Y_{-2}|\Ac_-,\Ac_+,\Ac_-\rangle\!\rangle\right)
\end{split}
\end{equation*}
and the string field is taken as
\begin{equation*}
\begin{split}
\Ac_+(z)=\int d^Dk u(k)c(z)e^{2ikX}(z),\quad
\Ac_-(z)=\int d^Dk \left(\tau(k)\eta
e^\varphi(z)+s_{\mu}(k)c\psi^{\mu}\right)e^{2ikX}(z).
\end{split}
\end{equation*}
After some tedious algebra one may get (integrating on the fly field $s_\mu$)
\begin{equation*}
\begin{split}
S_0=\frac1{g_o^2}\int d^Dxe^{-V_\mu x^\mu}&\left(\frac1{4\AP}u^2(x)-\frac12\pd_\mu\tau^2(x)+\frac1{4\AP}\tau^2+
\frac9{16}\left(\frac{4}{3\sqrt{3}}\right)^{\AP V^2}\tit^2(x)\tilde{u}(x)\right)
\end{split}
\end{equation*}
with
\begin{equation*}
\tilde{\varphi}(x)=e^{-\AP\left(\log\frac{4}{3\sqrt{3}}\right)\Box}\varphi(x).
\end{equation*}
Setting $V=0$ one restores the usual action coming from SFT. Also we notice that
\begin{equation*}
e^{+\AP\left(\log\frac{4}{3\sqrt{3}}\right)\Box}e^{-V_\mu x^\mu}=\left(\frac{4}{3\sqrt{3}}\right)^{\AP V^2}e^{-V_\mu x^\mu}.
\end{equation*}
This explains the form of interaction in (\ref{action_premodelo}). To make a connection with (\ref{action_premodelo}) one has to rescale fields and the coupling $g_o$ by appropriate powers of $\AP$.


\section{Derivation of $\Nc_{\mu\nu}$}

To vary an arbitrary (infinite) polynomial function of $\Box$ acting on a field $\phi$ w.r.t. the metric one finds useful the following relations
\begin{equation*}
\int d^Dx\sqrt{-g}A\delta\Box B=-\int d^Dx\sqrt{-g}\frac{\delta g^{\alpha\beta}}2\left(\pd_\alpha A\pd_\beta B+\pd_\beta A\pd_\alpha B-g_{\alpha\beta}\left(\pd_\mu A\pd^\mu B+A\Box B\right)\right),
\end{equation*}
\begin{equation*}
\int d^Dx\sqrt{-g}A\delta(\Box^n) B=\int d^Dx\sqrt{-g}\sum_{l=0}^{n-1}(\Box^l A)\delta\Box\Box^{n-1-l}B,
\end{equation*}
where we assume that boundary terms vanish.
If $\Gc(\Box)=\sum\limits_{n=0}^\infty g_n\Box^n$ then
\begin{equation*}
\int d^Dx\sqrt{-g}A\delta(\Gc) B=\sum_{n=1}^{\infty}g_n\int d^Dx\sqrt{-g}\sum_{l=0}^{n-1}(\Box^l A)\delta\Box\Box^{n-1-l}B.
\end{equation*}
Variation of the closed string tachyon potential w.r.t. the metric gives
\begin{equation*}
\begin{split}
{\Nc_c}_{\mu\nu}=\frac12\sum_{n=1}^{\infty}{g_c}_n{\AP}^n\sum_{l=0}^{l=n-1}
&\left(\pd_\mu\Box^l\left(e^{-\Phi}V'\right)\pd_\nu\Box^{n-l-1}T+
\pd_\nu\Box^l\left(e^{-\Phi}V'\right)\pd_\mu\Box^{n-l-1}T-\right.\\
&\left.-g_{\mu\nu}\left(g^{\rho\sigma}\pd_{\rho}\Box^l\left(e^{-\Phi}V'\right)\pd_\sigma\Box^{n-l-1}T+
\Box^l\left(e^{-\Phi}V'\right)\Box^{n-l}T\right)\right)
\end{split}
\end{equation*}
where we have used $\Gc_c=\sum\limits_{n=0}^{n=\infty}{g_c}_n{\AP}^n\Box^n$.
Variation of the open string tachyon potential w.r.t. the metric gives
\begin{equation*}
\begin{split}
{\Nc_o}_{\mu\nu}=\frac12\sum_{n=1}^{\infty}{g_o}_n{\AP}^n\sum_{l=0}^{l=n-1}
&\left(\pd_\mu\Box^l\left(\widehat{e^{-\Phi/2}}v'\right)\pd_\nu\Box^{n-l-1}\tau+
\pd_\nu\Box^l\left(\widehat{e^{-\Phi/2}}v'\right)\pd_\mu\Box^{n-l-1}\tau-\right.\\
&\left.-g_{\mu\nu}\left(g^{\rho\sigma}\pd_{\rho}\Box^l\left(\widehat{e^{-\Phi/2}}v'\right)\pd_\sigma\Box^{n-l-1}\tau+
\Box^l\left(\widehat{e^{-\Phi/2}}v'\right)\Box^{n-l}\tau\right)\right)
\end{split}
\end{equation*}
where we have used $\Gc_o=\sum\limits_{n=0}^{n=\infty}{g_o}_n{\AP}^n\Box^n$.
Variation of the $\widehat{e^{-\Phi/2}}$ term in front of the open string tachyon potential w.r.t. the metric gives
\begin{equation*}
\begin{split}
{\Nc_d}_{\mu\nu}=\frac12\sum_{n=1}^{\infty}{g_d}_n{\AP}^n\sum_{l=0}^{l=n-1}
&\left(\pd_\mu\Box^l\left(v\right)\pd_\nu\Box^{n-l-1}e^{-\Phi/2}+
\pd_\nu\Box^l\left(v\right)\pd_\mu\Box^{n-l-1}e^{-\Phi/2}-\right.\\
&\left.-g_{\mu\nu}\left(g^{\rho\sigma}\pd_{\rho}\Box^l\left(v\right)\pd_\sigma\Box^{n-l-1}e^{-\Phi/2}+
\Box^l\left(v\right)\Box^{n-l}e^{-\Phi/2}\right)\right)
\end{split}
\end{equation*}
where we have used $\Gc_d=\sum\limits_{n=0}^{n=\infty}{g_d}_n{\AP}^n\Box^n$. Non-local term $\Nc_{\mu\nu}$ which enters equation (\ref{EOM_gm}) is given by
\begin{equation}
\begin{split}
\Nc_{\mu\nu}=2{\Nc_c}_{\mu\nu}+\frac{4\kappa^2}{g_o^2}\left({\Nc_o}_{\mu\nu}+{\Nc_d}_{\mu\nu}\right).
\end{split}
\label{explicitNc}
\end{equation}
For simplicity and without any loss of generality we assume ${g_c}_0={g_o}_0={g_d}_0=1$ so that these non-local operators do not change a constant.


\end{document}